\documentclass[submission,copyright,creativecommons]{eptcs}
\providecommand{\event}{Linearity 2012}

\usepackage{graphics}
\usepackage{graphicx}
\usepackage{listings}
\usepackage{todonotes}
\usepackage{amsmath}
\usepackage{amsthm}
\usepackage{amssymb}
\usepackage{url}

\usepackage{pgf, tikz}
\usepackage{in_simple}
\title{Extending the Interaction Nets Calculus by Generic Rules}
\author{Eugen Jiresch\thanks{The author was supported by the Austrian
Academy of Sciences (\"OAW) under grant no.\ 22932 and the Vienna PhD School of
Informatics.} \\
\email{jiresch@logic.at}
\institute{Institute for Computer Languages \\
			Vienna University of Technology}}
\def\authorrunning{Eugen Jiresch}
\def\titlerunning{Extending the Interaction Nets Calculus by Generic Rules}

\newcommand{\lwint}{$\;\stackrel{int}{\longrightarrow}\;$}
\newcommand{\lwcom}{$\;\stackrel{com}{\longrightarrow}\;$}
\newcommand{\lwcol}{$\;\stackrel{col}{\longrightarrow}\;$}
\newcommand{\lwsub}{$\;\stackrel{sub}{\longrightarrow}\;$}
\newcommand{\mlwint}{\stackrel{int}{\longrightarrow}}
\newcommand{\mlwcom}{\stackrel{com}{\longrightarrow}}
\newcommand{\mlwcol}{\stackrel{col}{\longrightarrow}}
\newcommand{\mlwsub}{\stackrel{sub}{\longrightarrow}}

\begin{document}

\maketitle

\begin{abstract} 
We extend the textual calculus for interaction nets by generic rules and
propose constraints to preserve uniform confluence. Furthermore, we discuss the
implementation of generic rules in the language \emph{inets}, which is based on
the lightweight interaction nets calculus.
\end{abstract}

%%%%%%%%%%%%%%%%%%%%%%%%%%%%%%%%%%%%%%%%%%%%%
% INTRODUCTION
%%%%%%%%%%%%%%%%%%%%%%%%%%%%%%%%%%%%%%%%%%%%%
\section{Introduction and Overview}
\label{sec:intro}
% Models of computation are the basis for many programming languages,
% e.g., for reasoning on formal properties of programs such as 
% correctness and termination.
\emph{Interaction nets} are a model of computation based on graph
rewriting. Programs and data are represented as graphs (\emph{nets}), and execution of
a program is modeled by manipulating the net based on \emph{rewrite} or
\emph{reduction} \emph{rules}.
%\todo{better arguing why INs are nice (recent Ian paper?)}

The theory behind interaction
nets is well developed: they enjoy several useful properties
such as uniform confluence and locality of reduction: single computational
steps in a net do not interfere with each other, and thus may be performed in
parallel. Additionally, interaction nets share computations:
reducible expressions cannot be duplicated, which is beneficial
for efficiency in computations.
Furthermore, the graphical notation of interaction nets automatically provides
a visualization of an algorithm. Such a visualization can even show formal
properties of programs that might be hard to prove in a textual programming
language \cite{DBLP:conf/tamc/Mackie10}.

Our goal is to promote interaction nets to a practically usable
programming language. Unfortunately, the beneficial properties of interaction nets impose
strong restrictions on the shape of rules: this makes it hard to express
features such as higher-order functions or side effects. In this paper, we
improve this deficiency by extending the textual calculus for interaction nets
by \emph{generic rules}. In addition, we define constraints on these rules to
preserve uniform confluence, which is the basis for parallel evaluation.
Despite the merits of the graphical notation, the textual calculus for
interaction nets
\cite{DBLP:conf/ppdp/FernandezM99} is indispensable: it provides a precise
semantics for the mechanics of the graphical rewriting rules. Furthermore, it forms the basis for implementations
of interaction nets based languages \cite{DBLP:journals/eceasst/HassanMS10}.

% For these reasons, interaction nets
% provide a promising basis for a future programming language, much like the
% $\lambda$-calculus provides the basis for functional programming languages.

%Using ideas of \cite{DBLP:journals/eceasst/HassanMS10}, 
We complement our
previous work \cite{GTR_Techrep}, which defined generic rules in the graphical
setting of interaction nets. Defining generic rules in the textual calculus
gives a precise semantics to the graphical notation which we
introduced previously.
In addition, we describe the ongoing implementation of generic rules in the
interaction nets based language \emph{inets}. We show that the implementation
satisfies the constraints for generic rules, and hence preserves uniform
confluence. Our contributions can be summarized as follows:
\begin{itemize}
  \item We extend the interaction nets calculus with generic rules. In
  particular, we provide a precise definition of \emph{variadic}
  (arbitrary-arity) rules such as duplication and deletion.
  \item We describe the implementation of generic rules in the programming
  language \emph{inets}.
  \item We show that the generic rule constraints are satisfied in the calculus
  and the implementation, ensuring uniform confluence of reduction.
\end{itemize}
In the following section, we introduce interaction nets and the lightweight
calculus. Section \ref{sec:generic-rules-for-the-lighweight-calculus} defines
generic rules for the lightweight calculus. We discuss our ongoing
implementation of generic rules in \emph{inets} in Section
\ref{sec:implementation}. Finally, we conclude in Section \ref{sec:discussion}.

%%%%%%%%%%%%%%%%%%%%%%%%%%%%%%%%%%%%%%%%%%%%%
% PRELIMINARIES
%%%%%%%%%%%%%%%%%%%%%%%%%%%%%%%%%%%%%%%%%%%%%
\section{Preliminaries}
In this section, we recall the main notions of interaction nets and the
lightweight calculus. We discuss the uniform confluence property in both the graphical and
the textual formalism. Preserving this property is a key challenge in the
introduction of generic rules.

%%%%%%%%%%%%%%%%%%%%%%%%%%%%%%%%%%%%%%%%%%%%%%%%%%
\subsection{Interaction Nets}
\label{sec:interaction-nets}
Interaction nets were first introduced in
\cite{lafont_interaction_1990}.  
A \emph{net} is a graph consisting of \emph{agents} (labeled nodes)
and
 \emph{wires} (edges).
Each agent has a fixed number of \emph{ports}. Wires connect agents through these ports.
We say that an agent is or arity $n$ if it has $n+1$ ports:
every agent has exactly one \emph{principal} port
(denoted by the arrow), all other ports are called \emph{auxiliary} ports. 
%The number of auxiliary ports denotes the \emph{arity} of the agent. 
Intuitively, agent labels
denote data or function symbols. Computation is
modeled by rewriting the graph, which is based on \emph{interaction rules}.

 
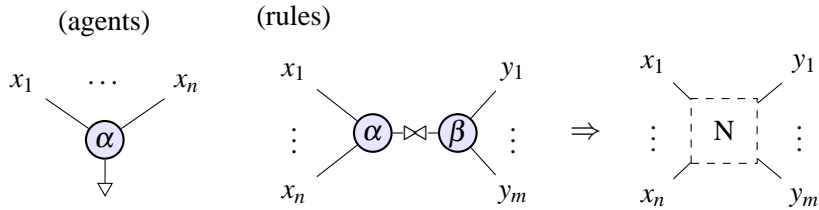
\begin{tikzpicture}
\matrix[column sep=0.5mm, row sep=2.5mm] {
	%1st
	&
	\node(label) {(agents)}; \\
	\node(x1) {$x_1$}; &
	\node(dot){\ldots}; &
	\node(xn) {$x_n$}; \\
	%2nd
	
	& \node[agent_small](a)	{$\alpha$}; & \\ \\
	%3rd
	
	& \node (p) {}; & \\
};
	\link{x1}{a}
	\link{xn}{a}
	\pp{a}{p}

\end{tikzpicture}
%basic IN rule schema
\begin{tikzpicture} 
\matrix[column sep=1.5mm, row sep=1.7mm] {
	%1st
	&
	\node(label) {(rules)}; \\
	&
	\node(x1) {$x_1$}; & & & & & &
	\node(y1) {$y_1$}; \\
	%2nd
	&
	\node(d) {\vdots};
	& \node[agent_small](a)	{$\alpha$}; & & & &
	\node[agent_small](b)		{$\beta$}; &
	\node(d2) {\vdots}; 
	\\
	%3rd
	&
	\node(xn) {$x_n$}; & & & & & &
	\node(yn) {$y_m$}; \\
};
	\activepair{a}{b}
	\link{a}{x1}
	\link{a}{xn}
	\link{b}{y1}
	\link{b}{yn}
	
	\node[right of=b, xshift=20pt] (is) {\large{$\Rightarrow$}};

\matrix[xshift=130pt, yshift=-10pt, column sep=1mm, row sep=2mm] {
	%0st
	\node(x1) {$x_1$}; & & & & &
	\node(y1) {$y_1$}; \\
	
	%1st
	\node(d) {\vdots};
	&
	& \node[shape=rectangle, draw= black, dashed, inner sep=3mm] (n) {N}; \\
	
	%2nd
	%0st
	\node(xn) {$x_n$}; & & & & &
	\node(yn) {$y_m$}; \\
};
	\link[]{n}{x1}
	\link[]{n}{xn}
	\link[]{n}{y1}
	\link[]{n}{yn}
	
	\node[right of=n](d2) {\vdots}; 

\end{tikzpicture}

 These rules apply to two 
nodes which are connected by
their \emph{principal ports}, forming an \emph{active pair}. We will refer to a set
of rules as \emph{interaction net system} (INS for short). 
% For example, the following
% INS models the addition of natural numbers (encoded by 0 and a successor
% function $S$):
% 
% \input{figures/addition}
%
This simple system allows for parallel evaluation of programs: 
if several rules are applicable at the same time, 
they can be applied in parallel without interfering
with each other. The main prerequisite for this parallelism is the \emph{uniform
confluence property} of the reduction relation induced by a set of rules.

\newtheorem{def:ucr}{Definition}[subsection]

\begin{def:ucr} [\textbf{Uniform Confluence}]
A relation $\rightarrow$ satisfies the uniform confluence property if the
following holds: if $N \rightarrow P$ and $N \rightarrow Q$ where
$P \neq Q$, then there exists some $R$ such that $P \rightarrow R \leftarrow
Q$.%
\footnote{Several publications on interaction nets, including
\cite{lafont_interaction_1990}, refer to this property as \emph{strong
confluence}. We use the term \emph{uniform confluence} (or $WCR\_1$ in the term
rewriting literature \cite{klop_term_1992,DBLP:journals/jfp/Niehren00}) in order to
account for the fact that if $P$ and $Q$ are distinct, then one step
is taken from either
net to reach a common reduct.}
\end{def:ucr}

\newtheorem{prop_ucr}[def:ucr]{Proposition}
\begin{prop_ucr} [Lafont \cite{lafont_interaction_1990}]
\label{prop:ucr}
Let $R$ be an interaction net system. The reduction relation $\Rightarrow$
induced by $R$ satisfies uniform confluence.
% Let $N$ be an interaction net. If $N \Rightarrow P$ and $N \Rightarrow Q$ where
% $P \neq Q$, then there exists a net $R$ such that $P \Rightarrow R \Leftarrow
% Q$. In this case the INS inducing $\Rightarrow$ is called \emph{uniformly confluent}.
% In the rewriting literature, strong confluence
% usually means the weaker property $\Rightarrow \cdot \Leftarrow \;\; \subseteq
% \;\; \Rightarrow^{\leq 1} \cdot \Leftarrow^{*} $. \todo{meaning of the
% formula?}

\end{prop_ucr}

Essentially, three properties of interaction net systems are sufficient for
uniform confluence \cite{lafont_interaction_1997}: \begin{description}%
\label{properties-lafont-for wcreins}
  \item[1) Linearity:] interaction rules cannot erase or duplicate ports.
  \item[2) Binary interaction:] agents can only be rewritten if they form an
    active pair, i.e., if they are connected via their principal ports.
  \item[3) No ambiguity:] 
    for each active pair $(S,T)$ of 
    agents there is \textit{at
      most one} rule 
      %whose LHS matches $S \sim T$.
      that can rewrite $(S,T)$. If $S$ and $T$ are the same agent, then
      rewriting $(S,T)$ must yield the same net as rewriting
      $(T,S)$.\footnote{See \cite{lafont_interaction_1997} for a detailed
      explanation of the idea behind the condition for $S=T$.}
\end{description}

\noindent
Later, we will see how generic rules influence these properties. Essentially, we
need to provide constraints on generic rules such that 3) is still satisfied.
%  
% In addition to parallel evaluation, reducible expressions (i.e., active pairs) 
% %in a program
% in a net
% cannot be duplicated: They are evaluated only once, which allows for sharing of
% computation.

%%%%%%%%%%%%%%%%%%%%%%%%%%%%%%%%%%%%%%%%%%%%%%%%%
\subsection{The Lightweight Interaction Calculus}
\label{sec:lightweight-calculus}

The lightweight calculus \cite{DBLP:journals/eceasst/HassanMS10} provides a
precise semantics for interaction nets. It handles application of rules as well
as rewiring and connecting of ports and agents. It uses the following ingredients:
\begin{description}
  \item[Symbols $\Sigma$] representing agents, denoted by $\alpha,\beta,\gamma$.
  \item[Names $\mathrm{N}$] representing ports, denoted by
  $x,y,z,x_1,y_1,z_1,\ldots$ . We denote sequences of names by
  $\overline{x},\overline{y},\overline{z}$.
  \item[Terms $\mathrm{T}$]	being either names or symbols with a number of
  subterms, corresponding to the agent's arity: $t = x \; | \; \alpha(t_1,\ldots,t_n)$ .
  $s,t,u$ denote terms,  $\overline{s},\overline{t},\overline{u}$ denote sequences of terms.
  \item[Equations $E$] denoted by $t=s$ where $t,s$ are terms, representing
  connections in a net. Note that $t = s$ is equivalent to $s = t$.
  $\Delta,\Theta$ denote multisets of equations.
  \item[Configurations $C$] representing a net by $\langle \overline{t}\;|\; \Delta
  \rangle$. $\overline{t}$ is the interface of the net, i.e., its ports that are not
  connected to an agent. All names in a configuration occur at most twice. Names
  that occur twice are called \emph{bound}.
  \item[Interaction Rules $\mathrm{R}$] denoted by $\alpha(\overline{x}) =
  \beta(\overline{y}) \longrightarrow \Theta$. $\alpha,\beta$ is the active pair of the
  left-hand side (LHS) of the rule and the set of equations $\Theta$ represents
  the right-hand side (RHS).
\end{description}
\newtheorem{def_no_ambiguity}[def:ucr]{Definition}
\noindent
The \emph{no ambiguity} constraint of Section \ref{sec:interaction-nets}
corresponds to the following definition for the lightweight calculus.

 \begin{def_no_ambiguity} [\textbf{No Ambiguity}]
We say that a set of interaction calculus rules $R$ is \emph{non-ambiguous} if
the following holds:
\begin{itemize}
  \item for all pairs of symbols $(\alpha, \beta)$, there is at most one rule 
  		$\alpha(\overline{x}) = \beta(\overline{y}) \longrightarrow \Theta \;\;$ or 
  		$ \beta(\overline{y}) = \alpha(\overline{x}) \longrightarrow \Theta \;\; \in R$.
  \item if an agent interacts with itself, i.e.,  $\alpha(\overline{x}) =
  \alpha(\overline{y}) \longrightarrow \Theta \in R$, then $\Theta$ equals $\Delta$
  (as multisets, modulo orientation of equations), where $\Delta$ is obtained
  from $\Theta$ by swapping all occurrences of $\overline{x}$ and $\overline{y}$.
\end{itemize}
\end{def_no_ambiguity}
\noindent
Rewriting a net is modeled by applying four \emph{reduction rules} to a
configuration with respect to a given set of interaction rules $R$:
\newtheorem{def:reduction_rules}[def:ucr]{Definition}

\begin{def:reduction_rules} [\textbf{Reduction Rules}]
\label{def:reduction_rules}
The four reduction rules of the lightweight calculus are defined as follows:
\begin{description}
  \item[Communication:] $\langle\; \overline{t} \;|\; x=t,x=u,\Delta \rangle
  \mlwcom \ \langle\; \overline{t} \;|\; t=u,\Delta \rangle $
  \item[Substitution:] $\langle\; \overline{t} \;|\; x=t,u=s,\Delta \rangle
  \mlwsub \ \langle\; \overline{t} \;|\; u[t/x]=s,\Delta \rangle $, where
  $u$ is not a name and $x$ occurs in $u$.
  \item[Collect] $\langle\; \overline{t} \;|\; x=t,\Delta \rangle
  \mlwcol \ \langle\; \overline{t}[t/x] \;|\; \Delta \rangle $, where $x$
  occurs in $\overline{t}$.
  \item[Interaction] $\langle\; \overline{t}  \;|\;  \alpha(\overline{t_1}) =
  \beta(\overline{t_2}),\Delta\rangle \mlwint \langle\; \overline{t} \;|\;
  \Theta',\Delta\rangle$, where $\alpha(\overline{x}) =
  \beta(\overline{y}) \longrightarrow \Theta \; \in \mathrm{R}$. $\Theta'$ denotes
  $\Theta$ where all bound names in $\Theta$ receive fresh names and
  $\overline{x},\overline{y}$ are replaced by $\overline{t_1}, \overline{t_2}$.
\end{description}
\end{def:reduction_rules}
\noindent
The reduction rules 
$\mlwcom$ and $\mlwsub$
replace names by terms:
this explicitly resolves connections between agents which are
generated by interaction rules.
 $\mlwcol$ also replaces
names, but only for the interface.
Naturally, $\mlwint$ models the application of interaction
rules: an equation corresponding to a LHS is replaced by the equations of the
RHS.

\newtheorem{exp:addition_calculus}[def:ucr]{Example}

\begin{exp:addition_calculus} 
The rules for addition of symbolic natural numbers
are expressed in the lightweight calculus as follows:
\begin{align} 
+(y,r) = S(x) & \;\;\longrightarrow\;\;  +(y, w) = x,\; r = S(w) \\
+(y,r) = Z  &  \;\;\longrightarrow\;\;  r = y
\end{align}
% The following reduction calculates $1 + 0$:
% \begin{align*}
%  	\langle\; r  \;|\;  +(r,0) = S(0)\;\rangle & \mlwint &
%  	\langle\; r  \;|\; r=S(x), +(x,0) = 0 \;\rangle \\
%  	&\mlwcol& \langle\; S(x)  \;|\;  +(x,0) = 0\;\rangle \\
%  	&\mlwint& \langle\; S(x)  \;|\;  x = 0 \;\rangle \\
%  	&\mlwcol& \langle\; S(0)  \;|\;  \;\rangle
% \end{align*}
\end{exp:addition_calculus}

\newtheorem{prop:ucr_calculus}[def:ucr]{Proposition}
\begin{prop:ucr_calculus} [\textbf{Uniform Confluence for the Lightweight
Calculus}]
\label{prop:ucr_calculus}
Let $\longrightarrow$ be the reduction relation induced by the four reduction
rules and a set of interaction rules $R$. If $R$ is non-ambiguous, then
$\longrightarrow$  satisfies uniform confluence.
\end{prop:ucr_calculus}
\begin{proof} [Proof (sketch)]
In
\cite{DBLP:conf/ppdp/FernandezM99}, uniform confluence is shown for the
interaction calculus, which is the predecessor of the lightweight calculus.
The main difference of the lightweight calculus to the previous one is that the
\emph{indirection} rule of the standard interaction calculus is now split into
\lwcom and \lwsub. However, this does not affect the property shown in
\cite{DBLP:conf/ppdp/FernandezM99}: all critical pairs (i.e., critical one-step
divergences in the reduction of a configuration) can be joined in one step.

\noindent
It is necessary that $R$ is non-ambiguous in order to prevent
non-determinism in the application of the \lwint rule, which could lead to non-joinable
divergences.
% In \cite{DBLP:conf/ppdp/FernandezM99}, uniform confluence of the interaction
% calculus, which is the basis for the lightweight calculus, is shown: all
% critical pairs between the four reduction rules can be joined in one step. For the
% additional rules of the lightweight calculus (communication and substitution),
% the same can be shown in a straightforward way. No ambiguity of interaction
% rules is required to prevent non-determinism in the reduction of a single
% equation with \lwint, which may lead to non-joinable critical pairs.
% \todo{better formulation: all rules commute (see [3]); no ambiguity is important
% s.t. int rules commute}
\end{proof}

%%%%%%%%%%%%%%%%%%%%%%%%%%%%%%%%%%%%%%%%%%%%%
% GENERIC RULES for the LW CALCULUS
%%%%%%%%%%%%%%%%%%%%%%%%%%%%%%%%%%%%%%%%%%%%%
\section{Generic Rules for the Lightweight Calculus}
\label{sec:generic-rules-for-the-lighweight-calculus}

In this section, we first introduce generic rules in the graphical setting of
interaction nets. Afterwards, we extend the lightweight interaction calculus in
order to express the semantics of generic rules.

\subsection{Generic Rules}
\label{sec:generic-rules}
% \todo[inline]{short intro and contribution: generic rules in calculus, maybe
% mention constraints}
Ordinary interaction rules describe the reduction of a pair of two concrete
agents (e.g., $0$ and $+$ in rule (1) above). \emph{Generic rules} allow one concrete
agent to interact with an arbitrary agent. This arbitrary, \emph{generic} agent
corresponds to a function variable, adding a higher-order character to
interaction nets. Such rules have already been used in several
publications (e.g.,
\cite{ian_mackie_yale:_1998}), usually to
model duplication and deletion of agents, albeit without a formal
definition of generic rules. 

\noindent
We distinguish two types of generic agents and rules based
on the arity of the agent:
\begin{description}
  \item[fixed generic agents] have a specific arity. They correspond to an
  arbitrary agent of exactly this number of ports. 
  \item[variadic agents] are of arbitrary arity. They correspond to any agent
  with any number of ports.
\end{description} 

\newtheorem{exp:delta_eps}{Example}[subsection]

\begin{exp:delta_eps} 
\label{exp:delta_eps}
The following rules model deletion and duplication via the agents
$\epsilon$ and $\delta$, where $\alpha$ is a \emph{variadic} agent.

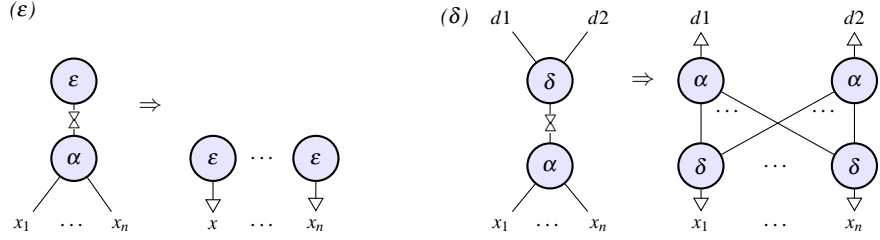
\tikzscale{0.75}
% \begin{tikzpicture}
% %nodes
%  \matrix[row sep=2mm, column sep=3mm] {
% 	%oth row
%     \node (ol1) 		 {\small{$d1$}}; &
%     \node (ol2) 		 {\small{$d2$}}; &
%     \node (or1)			 {\small{$d1$}}; &
%     \node (or2)			 {\small{$d2$}}; \\
%     %1st row
%     \node[agent] (delta) {$\delta$}; & &
%     \node (or)			 {}; &
%     \node (or)			 {}; \\
%     %2nd row
%     &
%     \node (is)		     {\large{$\Rightarrow$}}; &
% 	\node (ar1) [agent] {a}; &
% 	\node (ar2) [agent] {a}; \\
%     %3rd row
% 	\node (a) [agent] {a}; & \\
% 	\\
%  };
%  
%  %wires
%  \link{delta}{ol1}
%  \link{delta}{ol2}
%  \activepair{delta}{a}
%  
%  \pp{ar1}{or1}
%  \pp{ar2}{or2}
% 
% \end{tikzpicture}

%agents with 2 ports
\hspace{-10pt}
\begin{tikzpicture}
\matrix[row sep=4mm, column sep= 1mm] {
	%0
	\node (label)				{($\epsilon$)};
	\\
	&
	\node (eps)		[agent]		{$\epsilon$}; & \\
	%1
	&
	\node (b)		[agent]		{$\alpha$}; & & & & & & & &
	\node (er1)		[agent]		{$\epsilon$}; &
	\node (p1)					{\ldots}; &
	\node (er2)		[agent]		{$\epsilon$}; &
	 \\	
	%2
	\node (x)					{\small{$x_1$}}; &
	\node (p2)					{\ldots}; &
	\node (y)					{\small{$x_n$}}; & & & & & & &
	\node (xr)					{\small{$x$}}; &
	\node (p3)					{\ldots}; &
	\node (yr)					{\small{$x_n$}}; \\
};
	\node (is)		[right of=eps, yshift=-13pt, xshift=10pt]		
	{\large{$\Rightarrow$}};

	%wires
	\activepair{eps}{b}
	\link{b}{x}
	\link{b}{y}
	\pp{er1}{xr}
	\pp{er2}{yr}

\end{tikzpicture}
% \begin{tikzpicture} 
% 
% \matrix[row sep=5mm, column sep= 3mm] {
% 	%0
% 	\node (eps)		[agent]		{$\epsilon$}; \\
% 	%1
% 	\node (a)		[agent]		{a}; \\	
% };
% 	\node (is)		[right of=eps, yshift=-13pt]			{\large{$\Rightarrow$}};
% 
% 	%wires
% 	\activepair{eps}{a}
% 
% \end{tikzpicture}
\hspace{20pt}
\begin{tikzpicture}	
%nodes
 \matrix[row sep=3mm, column sep=1mm] {
	%oth row
	
	\node (label)				{($\delta$)}; &
    \node (ol1) 		 {\small{$d1$}}; & &
    \node (ol2) 		 {\small{$d2$}}; & & &
    \node (or1)			 {\small{$d1$}}; & &
    \node (or2)			 {\small{$d2$}}; \\
    %1st row
    & &
    \node[agent] (delta) {$\delta$}; & &
    \node[] (is)		     {\large{$\Rightarrow$}};  & &
	\node (ar1) [agent, label=290:\ldots] {$\alpha$}; &
	\node (p1)					{};  &
	\node (ar2) [agent, label=250:\ldots] {$\alpha$}; 
    %2nd row
    \\
    %3rd row &
	& &
	\node (a) [agent] {$\alpha$}; & & & &
    \node[agent] (dr1) {$\delta$}; &
	\node (p1)					{\ldots}; &
    \node[agent] (dr2) {$\delta$}; \\
	%4th row
	&
    \node (ol3) 		 {\small{$x_1$}}; &
	\node (p1)					{\ldots}; &
    \node (ol4) 		 {\small{$x_n$}}; & & &
    \node (or3)			 {\small{$x_1$}}; &
	\node (p1)					{\ldots}; &
    \node (or4)			 {\small{$x_n$}}; \\
 };
 
 %wires
 \link{delta}{ol1}
 \link{delta}{ol2}
 \activepair{delta}{a}
 \link{a}{ol3}
 \link{a}{ol4}
 
 \pp{ar1}{or1}
 \pp{ar2}{or2}
 \pp{dr1}{or3}
 \pp{dr2}{or4}
 \link{dr1}{ar1}
 \link{dr1}{ar2}
 \link{dr2}{ar1}
 \link{dr2}{ar2}

\end{tikzpicture}
\tikzscale{1}
\end{exp:delta_eps}

Informally, $\epsilon$ deletes any agent $\alpha$ and propagates itself to
$\alpha$'s ports, deleting connected agents in subsequent steps. Similarly,
$\delta$ duplicates an arbitrary agent and the net connected to it.

The dots at the ports of the variadic agent $\alpha$ indicate that its arity is
arbitrary, i.e., any active pair $(\delta, A)$ matches this rule (where $A$ may
be any agent). While this notation is intuitive, it does not give a precise
definition of the semantics of generic rule application. In particular, the RHS
of the $\delta$ rule has multiple sets of arbitrarily many ports and agents,
which may make it more difficult to comprehend. Hence, we provide
a definition of generic rule application in the lightweight calculus, clarifying
the mechanics that are associated with the graphical dot notation.
%  We present such a definition by extending the
%   \emph{lightweight calculus}, which is a textual representation of
%  interaction nets \cite{DBLP:journals/eceasst/HassanMS10}.

%\paragraph{\textbf{Generic Rules and Constraints}}
%%%%%%%%%%%%%%%%%%%%%%%%%%%%%%%%%%%%%%%%%%%%%%%%%%%%%%%%%%%%%%%%
\subsection{Fixed Generic Rules for the Lightweight Calculus}
We first extend the calculus by fixed generic rules. The more complex variadic
rules are defined in the following subsection.
Essentially, we introduce
additional symbols for generic agents. We then modify the
$\mlwint$ reduction rule to support generic agents.

\begin{description}
  \item[Generic Names $\mathrm{V}$] representing generic agents, denoted by
  $\phi,\psi,\rho$. Generic names may only occur in generic interaction rules.
  \item[Generic Rules $\mathrm{GR}$] denoted by $\alpha(\overline{x}) =
 \phi(\overline{y}) \longrightarrow \Theta$. $\Theta$ contains no generic
 names other than $\phi$.
\end{description}

\noindent
The reduction rule for interaction is extended to support matching and
application of generic rules.
\newtheorem{def:generic_interaction}{Definition}[subsection]

\begin{def:generic_interaction} [\textbf{Generic Interaction}]
\label{def:generic_interaction}
  $\langle\; \overline{t}  \;|\;  \alpha(\overline{t_1}) =
  \beta(\overline{t_2}),\Delta\rangle \mlwint \langle\; \overline{t} \;|\;
  \Theta',\Delta\rangle$, where $\alpha(\overline{x}) =
  \beta(\overline{y}) \longrightarrow \Theta \; \in \mathrm{R}$ or 
  $\alpha(\overline{x}) =
  \phi(\overline{y}) \longrightarrow \Theta \; \in \mathrm{GR}$ if $\beta$ and $\phi$
  have the same \emph{arity} (number of ports).
  %\todo{beta/phi too?} 
  In the latter
  case, $\Theta'$ equals $\Theta$ where all occurrences of $\phi$ are replaced
  by $\beta$ (in addition to using fresh names and replacing $\overline{x},\overline{y}$).
\end{def:generic_interaction}

\noindent
The above definition gives a precise semantics for the application of generic
rules with generic agents of fixed arity. Our approach is extended to generic
rules with variadic agents in Section \ref{sec-variadic-rules-for-the-lw-calculus}.

Note that the definition of generic interaction only modifies the behaviour of
the \lwint rule. The other three reduction rules are not affected by this change: they only
operate on configurations, which do not feature generic names.

%%%%%%%%%%%%%%%%%%%%%%%%%%%%%%%%%%%%%%
\subsection{Generic Rule Constraints}
\label{sec-generic-rule-constraints}
%Generic rules are substantially more powerful than ordinary interaction rules.
Unfortunately, generic rules introduce \emph{ambiguity} or \emph{overlaps} to
rule application: one equation could possibly be reduced by more than one
interaction rule. 
As mentioned in Proposition \ref{prop:ucr_calculus}, \emph{no ambiguity} is one of the
required properties for uniform confluence. Hence, overlaps may destroy the
nice properties of interaction nets (including parallel evaluation).
Therefore, overlaps caused by generic rules need to be prevented.

In \cite{GTR_Techrep}, we defined generic rule constraints to preserve
uniform confluence in the graphical setting of interaction nets. 
These
constraints can be translated to the lightweight calculus in a straightforward
manner. The Default Priority Constraint (DPC) corresponds to a modification of the
$\mlwint$ reduction rule, just as it restricts the reduction
relation in the graphical setting in \cite{GTR_Techrep}.

\begin{description}
  \item[Default Priority Constraint (DPC)] An equation $\alpha(\overline{t_1}) =
  \beta(\overline{t_2})$ can only be reduced using a generic rule if no matching
  ordinary rule exists, i.e., if $\alpha(\overline{x}) =
  \beta(\overline{y}) \longrightarrow \Theta \; \notin \mathrm{R}$.
  \item[Generic Rule Constraint (GRC)] If there is more than one generic
  rule that can be applied to a given equation $\alpha(\overline{t_1}) =
  \beta(\overline{t_2})$, there must exist an ordinary rule that can be applied
  as well.
\end{description}
The DPC restricts the behavior of $\mlwint$:
ordinary rules always have priority over generic rules. The GRC restricts the
set of generic rules $GR$. The combination of these
constraints prevents overlaps:

\newtheorem{prop:generic_rule_constraints}{Proposition}[subsection]

\begin{prop:generic_rule_constraints} 
\label{prop:generic_rule_constraints}
Let $R$ be a set of interaction rules (including generic rules) that satisfies
the GRC. If \lwint satisfies the DPC, then there is at most one rule that can
reduce an arbitrary equation  $\alpha(\overline{s}) =
  \beta(\overline{u})$.
\end{prop:generic_rule_constraints}
\begin{proof}
We distinguish two possible cases of overlaps:
\begin{enumerate}
  \item One ordinary and one generic rule can be applied to the same
equation (as defined in Definition \ref{def:generic_interaction}). Then, the
ordinary rule is chosen due to the DPC. 
  \item 	There are two generic rules that can be applied to the
	same equation. Then, by the GRC there must also be an ordinary equation that
	can be applied. This rule is again prioritized by the DPC.
\end{enumerate}
In both cases, there is only one possible rule that can be applied.
As with ordinary interaction rules, the
case of two ordinary rules with the same active pair is ruled out.
\end{proof}
With the DPC, a generic rule corresponds to a set of non-ambiguous ordinary
rules. The GRC eliminates a few obvious cases of rule overlaps.
Analogously to \cite{GTR_Techrep}(Proposition 3.3.3), we can now show uniform
confluence of the lightweight calculus with generic rules.

\newtheorem{prop_generic_ucr}[prop:generic_rule_constraints]{Proposition}
\begin{prop_generic_ucr}[\textbf{Uniform Confluence} ] 
Let $\longrightarrow$ be the reduction relation induced by the four reduction
rules and a set of interaction rules (including generic rules) $R$ 
%with generic rules 
that satisfies the GRC. If \lwint satisfies the
DPC, then $\longrightarrow$ satisfies the uniform confluence property.
\end{prop_generic_ucr}
\begin{proof} [Proof (sketch)]
The main argument is similar to the one used in Proposition
\ref{prop:ucr_calculus}: all critical pairs can be joined in one step. Generic
interaction rules do not affect the reduction rules \lwcol, \lwcom, \lwsub. Proposition \ref{prop:generic_rule_constraints} shows that the generic rule constraints
prevent any ambiguity that might arise from the application of the \lwint rule.
\end{proof}

%%%%%%%%%%%%%%%%%%%%%%%%%%%%%%%%%%%%%%%%%%%%%%%%%%%%%%%%%%%%%%%%%
\subsection{Variadic rules for the lightweight calculus}
\label{sec-variadic-rules-for-the-lw-calculus}
%%%%%%%%%%%%%%%%%%%%%%%%%%%%%%%%%%%%%%%%%%%%%%%%%%%%%%%%%%%%%%%%%
We now extend the lightweight calculus with variadic rules.
First, we define additional symbols to denote variadic agents and rules. We
exploit the fact that all ports of a variadic agent are handled in the same,
\emph{uniform} way when applying a rule.

Clearly, the lightweight calculus needs to capture the feature of arbitrary
arity (visualized by the dot notation) in a precise way. Intuitively, arbitrary
arity boils down to two mechanisms, as can be seen in the variadic rules for $\delta/\epsilon$ in
Example \ref{exp:delta_eps}:
\begin{enumerate}
  \item A single agent may have arbitrarily many
ports connected to it, like $\alpha$ in the LHS of both rules.
\item A net may be
connected to each of the arbitrarily many ports, resulting in arbitrarily many
agents. This can be seen in the RHS of the $\epsilon$-rule, which contains one
epsilon for each port.
\end{enumerate}   Note that in case 2), the net connected to each port is
the same, i.e., the ports are handled \emph{uniformly}.

These two aspects of variadic rules are captured in the notions of \emph{variadic
ranges} and \emph{names}.

\begin{description}
  \item[Variadic Ranges] denoted by $[x],[y],[z],\ldots$, where $x,y,z$ are
  names. A variadic range corresponds to the set of (arbitrarily many) ports of
  a variadic agent.
  \item[Variadic Names] \emph{VN} denoted by $x',y',z',\ldots$ . $x'$ denotes an
  arbitrary single port of the variadic range $[x]$. A variadic name may only
  appear in the RHS of a rule.
  \item[Variadic Rules] \emph{VR} are generic rules $\alpha(\overline{y}) =
 \phi([x]) \longrightarrow \Theta$ where $\Theta$ may contain:
 \begin{itemize}
  \item ordinary equations
  \item equations with variadic ranges
  \item equations with variadic names
\end{itemize}
 An equation in $\Theta$ must not have a variadic range and a
 variadic name at the same time.
\end{description}
Intuitively, ranges denote
the full set of ports of a variadic agent. Variadic names refer to a single port
of the variadic agent. All ports of a variadic agent are handled in the same way
when applying a rule. Therefore, an equation containing a variadic name
specifies how to treat each individual port of the variadic range. As with
regular names, variadic ranges and names may appear at most twice in a rule RHS.

Variadic rules capture the two mechanisms mentioned above: the
manipulation of a single, arbitrary port and the manipulation of the set of all
ports. These two operations are sufficient to provide the expressive power of variadic rules in the graphical
setting.

The application of
the $\mlwint$ reduction rule gets a bit more
complicated in the presence of variadic rules: we have to take the arity of the
agent corresponding to the varidiac agent into account when replacing an equation.
\newtheorem{def_variadic_interaction}[prop:generic_rule_constraints]{Definition}
\begin{def_variadic_interaction} [\textbf{Variadic Interaction}]
\label{def:variadic_interaction}
  A variadic interaction step is defined as $\langle\; \overline{r}  \;|\; 
  \alpha(\overline{t_\alpha}) = \beta(\overline{u}),\Delta\rangle \mlwint
  \langle\; \overline{r} \;|\; \Theta',\Delta\rangle$, where
  %$\alpha(\overline{s_1}) =
  %\beta(\overline{u_2}) \longrightarrow \Theta \; \in \mathrm{R}$ or 
  $\alpha(\overline{z}) =
  \phi([x]) \longrightarrow \Theta \; \in \mathrm{VR}$ and $\Theta'$ is
  instantiated from $\Theta$ as follows: (let $arity(\beta)=n$)
	\begin{itemize}
	  	\item if $t = \gamma([x]) \in \Theta$, then $t = \gamma(\overline{u}) \in
	  	\Theta'$.
	  	\item if $t = \gamma([y]) \in \Theta$ with $y \neq x$, then $t =
	  	\gamma(y_1,\ldots,y_n) \in \Theta'$.
	  	\item $t = s \in \Theta$ such that $t$ and $s$ contain one or more
	  	variadic names $x',y',\ldots$. We then add $n$ equations
	  	$t_1 = s_1, \ldots, t_n = s_n$ to $\Theta'$: $t_i = s_i$ ($1 \leq i \leq n$)
	  	equals $t = s$ where all occurrences of a variadic name $x'$ are replaced by
	  	$u_i$ (where  $\overline{u}=u_1,\ldots,u_n$) if the range $[x]$ occurs in
	  	the LHS or by the name $x_i$ otherwise.
% 	  	\item if $t = x' \in \Theta$, then ${t = u_1, \ldots,t = u_n} \in \Theta'$,
% 	  	where $\overline{u}=u_1,\ldots,u_n$.
% 	  	\item if $t = y' \in \Theta$ with $y \neq x$, then ${t = y_1, \ldots,t =
% 	  	y_n} \in \Theta'$.
	  	\item equations without variadic ranges or names are added to
	  	$\Theta'$ without change.
	  	\item all occurrences of $\phi$ in $\Theta$ are
	  	replaced by $\beta$ and all occurrences of $\overline{z}$ by
	  	$\overline{t_\alpha}$.
	\end{itemize}
\end{def_variadic_interaction}

Informally, a variadic range in $\Theta$ is replaced by $\beta$'s arguments
if it occurs in the rule LHS: for example, $[x]$ is replaced by $\overline{u}$ 
in the above definition. Otherwise, the range is replaced by $n$ fresh names,
where $n=arity(\beta)$. An equation with a variadic name is copied to $\Theta'$
$n$ times: if a variadic name corresponds to the variadic range in the LHS, it is
replaced by one of $\beta$'s arguments in each copy. Otherwise, it is replaced
by one of the fresh names of the corresponding variadic range: for example,
$t=y'$ is replaced by ${t = y_1, \ldots,t = y_n}$, where the names
$y_1,\ldots,y_n$ are the result of instantiating the range $[y]$.

\newtheorem{exp-variadic1}[prop:generic_rule_constraints]{Example}
\label{exp-variadic1}
\begin{exp-variadic1}
The variadic rule for deletion via the agent $\epsilon$ can be defined as
follows: $\epsilon = \phi([x]) \;\; \longrightarrow \;\;  \epsilon = x'$. Applying the
rule to the equation $\epsilon = A(u,v,t)$ yields $\{\epsilon = u, \epsilon = v, \epsilon =
t\}$: the name $x$ occurs in the LHS (via $[x]$), hence $e=x'$ is instantiated
three times for each of $A$'s arguments $u,v,t$.
\end{exp-variadic1}

\newtheorem{exp-variadic2}[prop:generic_rule_constraints]{Example}
\begin{exp-variadic2}
The variadic rule for duplication via the agent $\delta$ can be defined as
follows: $\delta(d_1,d_2) = \phi([x]) \;\; \longrightarrow \;\; d_1 = \phi([y]), \; d_2 = \phi([z]),
\; x' = \delta(y',z')$. Applying the rule to
the equation $\delta = A(u,v,t)$ yields 
$\{ d_1 = A(y_1,y_2,y_3),\; d_2 = A(z_1,z_2,z_3),\; u=\delta(y_1,z_1),\;
v=\delta(y_2,z_2),\; t=\delta(y_3,z_3) \}$: replacing the RHS-only variadic
ranges $[y],[z]$ yields the fresh names $y_1,y_2,y_3,z_1,z_2,z_3$, which are
used in the 3 instances of the equation $x' = \delta(y', z')$.
\end{exp-variadic2}

These examples show that the textual definition of variadic rules is very
concise. It clearly expresses the semantics of variadic rules in the graphical
setting (see Example \ref{exp:delta_eps}). Variadic ranges and names formalize
the mechanics expressed by the graphical dot notation.
% These examples show that with the above extensions to the lightweight calculus,
% generic rules can be expressed just as conveniently as in the graphical setting.
%\todo{more arguing why variadic rule def is important}

\subsection{Variadic rules with non-uniform port handling}
\label{sec-variadic-rules-with-nonuniform-port-handling}
The main characteristic of variadic rules is that every port of a variadic agent
is handled in the same way, i.e., connected to the same agent or identical nets.
This is strongly connected to the notion of arbitrarily many ports, making them
in a sense indistinguishable. For fixed generic agents, we can of course
distinguish between their ports and handle them in different ways during rule
application.

Both of these aspects of generic rules can be combined in the form of
\emph{non-uniform port-handling} \cite{GTR_Techrep}. In addition to their
arbitrarily many ports, variadic agents may have a fixed, finite number of
ports which may be handled specifically, or non-uniformly. Such a variadic agent
matches all agents with an arity greater or equal to the number of fixed ports.

This feature translates well to the lightweight calculus. Handling the fixed
ports of the variadic agent is expressed with ordinary equations.
% The idea of non-uniform port handling presented in \cite{GTR_Techrep}
% translates well to the lightweight calculus. 
Definition \ref{def:variadic_interaction} in the previous subsection states that only
equations with variadic ranges or names are treated in a special way. Ordinary
equations are handled the same way as in the ordinary $\mlwint$ rule. 
This means that the fixed ports are independent of the set of arbitrarily many
ports.
 As
an example, we recall the rules of the \emph{Maybe} monad from
\cite{GTR_Techrep}.
\lstset{% 
   basicstyle=\ttfamily,
   mathescape=true,
   language=Haskell,
   showstringspaces=false}
\newtheorem{exp_maybe_nonuniform}{Example}[subsection]
\begin{exp_maybe_nonuniform} 
\label{exp_maybe_nonuniform}
The Maybe monad is used in Haskell to model exception
handling. It is defined as follows:
\begin{lstlisting}
      data Maybe a   = Just a | Nothing
(1)   return x       = Just x
(2)   (Just x) >>= f = f x
(3)   Nothing  >>= f = Nothing
\end{lstlisting}
% The following interaction rules model the Haskell function definition 
% \lstinline;Nothing >>= f = Nothing; :
%The following set of rules models the \emph{Maybe} monad in interaction nets:
% 
% \tikzscale{0.75}
% \input{figures/maybe_rules_rev}
% \tikzscale{1}

\noindent
In the lightweight calculus, the Maybe monad is expressed by these rules
($\epsilon$ is defined in Example \ref{exp-variadic1}):
\begin{align*} 
Ret(r) = \phi([x]) &\; \longrightarrow \; \{ r = Jst(\phi([x])) \} & (1) \\
Jst(a) = \; \tiny{>>=}(b) &\; \longrightarrow \; \{ a = b \}  & (2) \\ 
No = \; \tiny{>>=}(b) &\; \longrightarrow \; \{ Aux = b \}  & (3a)\\
Aux = \phi(r,[x]) &\; \longrightarrow \; \{ \epsilon = x', No = r \} & (3b) \\
Aux = ret(r) &\; \longrightarrow \; \{ No = r \} & (GRC)
\end{align*}

The rules are labeled in correspondence to the lines of Haskell's Maybe monad
definition. The (GRC) rule is added to satisfy the generic rule constraint,
eliminating ambiguity between rules (1) and (3b).
In rule (3b), $\phi$ has both a variadic range $[x]$ and a single port $r$ which
is handled non-uniformly.
Just like $f$ in the Haskell definition, this rule accepts an arbitrary
function, including partially applied (curried) functions, e.g.,
\lstinline;(+1);.
\end{exp_maybe_nonuniform}

\newtheorem{exp_select}[exp_maybe_nonuniform]{Example}

\begin{exp_select} 
Consider a function \lstinline;pick;, which picks the nth element of a list or
returns \lstinline;Nothing; if that element does not exist:
\begin{lstlisting} 
pick :: Int -> [a] -> Maybe a
pick n []    = Nothing
pick 0 (x:xs) = Just x
pick n (x:xs) = pick (n-1) xs
\end{lstlisting}
The corresponding interaction rules can be defined as follows (the arguments
are swapped for better readability of the rules):
\begin{align*} 
pick(r, n) = Nil &\; \longrightarrow \; \{ r = No, \epsilon = n \} & (1) \\
pick(r, n) = Cons(x,xs) \; &\; \longrightarrow \; \{ pickH(r,x,xs) = n \}  & (2)
\\ 
pickH(r,x,xs) = Z &\; \longrightarrow \; \{ r=Jst(x), \epsilon=xs \}  & (3)\\
pickH(r,x,xs) = S(n) &\; \longrightarrow \; \{ pick(r,n) = xs, \epsilon = x  \}
& (4)
\end{align*}

\noindent
Using the rules for the \emph{Maybe} monad from the previous example, we can
evaluate the expression \\ \lstinline;Nothing >>= (pick 0); :
\begin{align*}
 	\langle\; r  \;|\;  No = \; \tiny{>>=}(f), f = pick(r,Z) \;\rangle 
 	\;\;\mlwint
 	 \langle\; r  \;|\; Aux = f, f = pick(r, Z) \;\rangle
 	\;\mlwcom \langle\; r  \;|\;  Aux = pick(r,Z) \;\rangle \\ 
 	\;\mlwint \langle\; r  \;|\;  No = r, \epsilon= Z \;\rangle 
 	\;\mlwint \langle\; r  \;|\;  No = r \;\rangle 
 	\;\mlwcol \langle\; No  \;|\;  \;\rangle 
\end{align*}
% \begin{align*}
%  	\langle\; r  \;|\;  No = \; \tiny{>>=}(f), f = pick(r,0) \;\rangle & \mlwint
%  	& \langle\; r  \;|\; Aux = f, f = pick(r, 0) \;\rangle \\
%  	&\mlwcom& \langle\; r  \;|\;  Aux = pick(r,0) \;\rangle \\
%  	&\mlwint& \langle\; r  \;|\;  No = r, \epsilon= 0 \;\rangle \\
%  	&\mlwint& \langle\; r  \;|\;  No = r \;\rangle \\
%  	&\mlwcol& \langle\; No  \;|\;  \;\rangle \\
% \end{align*}
\end{exp_select}

We conclude this section with an example on how to use generic rules to model
higher-order functions. We describe the well-known \texttt{map} function, which
is modeled in a similar way to the monadic operator $>>=$.
\newtheorem{exp_map}[exp_maybe_nonuniform]{Example}
\begin{exp_map}
The function \emph{map} is defined as follows:
\begin{lstlisting}
map :: (a -> b) -> [a] -> [b]
map f [] = []
map f (x:xs) = (f x):(map f xs)
\end{lstlisting}
Using variadic rules with non-uniform port handling, we can define
\emph{map} in the lightweight calculus. Since the \texttt{f} is deleted
and duplicated in the definition above, we again use $\epsilon$ and $\delta$
agents.
\begin{align*} 
map(r) = Nil &\; \longrightarrow \; \{ mapN = r \} & (1) \\
map(r) = Cons(a,as) \; &\; \longrightarrow \; \{ mapC(a,as) = r \}  & (2)
\\ 
mapN = \phi(r,[x]) &\; \longrightarrow \; \{ Nil = r, \epsilon = x' \}  & (3)\\
mapC(a, as) = \phi(r,[x]) &\; \longrightarrow \; \{ Cons(s,t)=
r, \phi(s,[y]) = a, \phi(t,[z])= u, \\
& \;\;\;\;\;\;\;\;\;\;\;\;\;\; map(u) = as, \delta(y', z') = x'  ) \} & (4)
\end{align*}
\end{exp_map}

%%%%%%%%%%%%%%%%%%%%%%%%%%%%%%%%%%%%%%%%%%%%%
% IMPLEMENTATION
%%%%%%%%%%%%%%%%%%%%%%%%%%%%%%%%%%%%%%%%%%%%%
\section{Implementation}
\label{sec:implementation}
In this section, we discuss the ongoing implementation of generic rules in the
prototype language \emph{inets}\cite{inets_project_site}, which is based on the
interaction nets calculus.
%The abstract machine evaluating nets is based on the interaction nets calculus.
We can show that our implementation satisfies the generic rule constraints
defined in Section \ref{sec-generic-rule-constraints}.
We will only describe the implementation of fixed generic rules, as the
implementation of variadic rules is still work in progress.

%\todo{explain why new syntax is introduced (inets!)}
\emph{inets} consists of two components: the \emph{inets} language and compiler,
and the runtime system. The \emph{inets} language is based on the interaction
calculus, and is compiled to C code, which is executed by the runtime system. 
The runtime
holds data structures for managing interaction rules as well as agents and
connections between them. For example, the interaction rules for addition of
natural numbers is implemented by the following piece of code (We use
the syntax of the \emph{inets} language, where equations are denoted by
\lstinline;><; on the LHS, and by $\sim$ on the RHS):
\begin{lstlisting} 
Add(r, y) >< Z  =>  r$\sim$z;
Add(r, y) >< S(x)  =>  r$\sim$S(w), x$\sim$Add(w, y);
\end{lstlisting}

The implementation of generic rules allows us to define interaction net systems
similar to the Maybe monad in Example
\ref{exp_maybe_nonuniform} (the keyword
\lstinline;ANY; denotes a generic agent):
\begin{lstlisting} 
Return(r) >< ANY(x)  => r$\sim$Just(ANY(x))
Bind(r)   >< Just(x) => r$\sim$x;
Bind(r)   >< Nothing => r$\sim$Aux;
Aux 	  >< ANY(r)  => r$\sim$Nothing;
\end{lstlisting}

 For a complete and detailed description of \emph{inets} and
its runtime, we refer to \cite{DBLP:journals/entcs/HassanMS09}. Here, we will concentrate on the
extension of the matching function to support generic rules. 

\subsection{Generic Rule Constraints}
Individual
interaction rules (both ordinary and generic) are represented as C functions
that take references of the two agents of an active pair as arguments. These
functions replace an active pair by the corresponding RHS net and connect it to
the rest of the net accordingly. The runtime maintains a table that maps a pair
of agent symbols to an interaction rule function. This table also
contains entries for fixed generic rules, with a special symbol for generic
agents of a specific arity. The following pseudocode describes the matching and
reduction function, where $\phi_n$ denotes the generic agent of arity $n$:
\lstset{language = C}
\begin{lstlisting} 
void reduce(agent1, agent2) {
  // is there an ordinary rule for the active pair?
  rulePtr rule = ruleTable[agent1][agent2]
  if (rulePtr == null) {
      // is there a fixed generic rule matching the pair?
      bool success = reduceGeneric(agent1, agent2)
      if (!success) 
        error("no matching rule!")
  }
  else {
    rule(agent1, agent2) //apply the ordinary rule
  }
}

bool reduceGeneric(agent1, agent2) {
  //is there a fixed generic rule with matching arity?
  n = arity(agent1)
  m = arity(agent2)
  rulePtr rule = ruleTable[$\phi_n$][agent2]
  if (rule == 0) {
    rule = ruleTable[$\phi_m$][agent1]
    if (rule == 0)
      return false  // no matching generic rule
  }
  // apply the generic rule
  rule(agent1, agent2)
  return true
}

\end{lstlisting}

The Generic Rule Constraint (GRC) is a property of the set of interaction rules,
and can thus be verified at compile time. We check each generic rule for
overlaps with already compiled generic rules, as shown by the following
pseudocode:

\begin{lstlisting} 

void checkGRC(Rule r) {
  if (r is a generic rule) {
    let A be the ordinary agent of r$'$s LHS
    let $\phi_m$ be the generic agent of r$'$s LHS
    n = arity(A)
    if (a generic rule with LHS B >< $\phi_n$ exists 
     and arity(B) = m ) {
      // we have two overlapping generic rules
      if (no ordinary rule with LHS A >< B exists)
      	error("generic rule overlap!")
    }
    add r to the existing rules
  }
}

\end{lstlisting}

Clearly, the implementation needs to satisfy the generic rule constraints of
Section \ref{sec-generic-rule-constraints}. Otherwise, multiple generic rules
may overlap, resulting in non-determinism in the evaluation of the program. It
is straightforward to see that the pseudocode above satisfies the generic rule
constraints:

\newtheorem{prop:inets_generic_rule_constraints}{Proposition}[subsection]
\begin{prop:inets_generic_rule_constraints}
\label{prop:inets_generic_rule_constraints}
The implementation of generic interaction rules in \emph{inets} satisfies the DPC and
GRC.
\end{prop:inets_generic_rule_constraints}
\begin{proof}
Consider the function \lstinline;reduce;. The application of a generic rule via
\\ \lstinline;reduceGeneric; is only attempted if no ordinary interaction rule
exists. Hence, the DPC is satisfied. For the GRC, consider \lstinline;checkGRC;:
if the generic rule currently being checked overlaps with a previous generic
rule and no matching ordinary rule exists (i.e., the GRC is violated), an error
is reported.
\end{proof}

\paragraph{\textbf{Implementation of Variadic Rules}}
% The implementation of variadic rules in \emph{inets} is currently work in
% progress. In principle, matching of variadic rules can be done in a
% way similar to the fixed case presented above. In addition, we need to provide
% constructs for variadic names and ranges in the \emph{inets} language.
% \todo[inline]{a bit more info, since this has progressed}
The current version of \emph{inets} also supports variadic rules. During
compilation, a variadic rule is translated into a set of fixed generic rules,
one for each possible arity. While variadic rules are theoretically defined for
agents with arbitrarily large arities, we can easily identify the maximum arity
$n$ of all agents in the current program. Hence, we only add $n$ fixed versions
of the variadic rule. After this step, the set of rules only contains fixed
generic rules, which are handled as described above.

%%%%%%%%%%%%%%%%%%%%%%%%%%%%%%%%%%%%%%%%%%%%%
% DISCUSSION
%%%%%%%%%%%%%%%%%%%%%%%%%%%%%%%%%%%%%%%%%%%%%
\section{Discussion}
\label{sec:discussion}

\subsection{Related Work}
The textual calculus for interaction nets was initially defined in
\cite{DBLP:conf/ppdp/FernandezM99}. We based our extensions on the improved
\emph{lightweight} calculus, which was introduced in
\cite{DBLP:journals/eceasst/HassanMS10}. A different approach to higher-order
computation in the interaction calculus can be found in
\cite{DBLP:journals/entcs/FernandezMP07}.

Besides \emph{inets}, several implementations of interaction nets evaluators
exist. Examples are \emph{amineLight} \cite{DBLP:journals/eceasst/HassanMS10} or
INblobs \cite{jose_bacelar_almeida_tool_2008}.
To the best of our knowledge, none of these systems support generic interaction
rules. Another recent tool is PORGY
\cite{DBLP:journals/corr/abs-1102-2654}, which can be used to analyse and
evaluate interaction net systems with a focus on evaluation strategies.

The extension of interaction nets to a practically usable programming
language has been the topic of several publications. For example, in
\cite{DBLP:journals/eceasst/MackiePV07} the authors propose a way to
represent higher-order recursive functions like \emph{fold} or \emph{unfold}.
\emph{Nested patterns}, an extension to allow more complex interaction rules,
have been dealt with in
% Besides generic rules, we have also been involved in the
%implementation of nested patterns, an extension to allow more complex
\cite{DBLP:journals/corr/abs-1003-4562,DBLP:journals/entcs/HassanS08}. They
combine well with generic rules.

\subsection{Conclusion and Further Work}
 In this paper, we extended the interaction nets calculus by generic rules.
Our previous work on generic rules \cite{GTR_Techrep} did not consider this
textual calculus. Instead, we defined generic rules and their constraints
(including a basic type system) in the graphical setting.
The
extension of the lightweight calculus provides an alternative precise semantics
to the graphical notation of generic interaction rules. This is particularly
important for variadic rules, which use a dot notation to express an arbitrary
number of ports. Our approach using variadic names and ranges is concise 
%and offers the same expressive power as 
and precisely formulates the mechanics of
the dot notation of the graphical
rewrite rules.

In addition, we discussed the ongoing implementation of generic rules in
\emph{inets}. This implementation satisfies the generic rule constraints DPC and
GRC and hence preserves uniform confluence.

%\todo{generic rules with arbitrary arity conclusion}
Generic rules 
%are a powerful tool 
%that 
allow us
to conveniently express higher-order functions. An example can be found in
\cite{ICGT_full_ECEASST}, where an interaction nets encoding of \emph{map} is
given. Generic agents assume a role similar to function variables in functional
programming. Hence the definition of higher-order functions via interaction rules closely mimics functional programs without the need for explicit lambda and function application agents. This
brings interaction nets closer to a practically usable programming language.

%. For example, 
A major motivation for formally dealing with generic rules comes from our own
previous work \cite{ICGT_full_ECEASST,GTR_Techrep}, where we used generic rules
to model side effects in interaction nets via monads. As part of future work, we
plan to define an abstract, unified interface for monads, similar to type
classes in Haskell. The \emph{agent archetype} approach of
\cite{DBLP:journals/eceasst/MackiePV07} may be a possible direction to achieve
this.
% In addition, we are involved in the implementation of an interaction nets based
% programming language \cite{inets_project_site}, and already extended it by
% generic rules. 
In addition, we will continue to contribute to \emph{inets}. Moreover, we are
currently investigating an implementation of interaction nets on parallel
hardware (GPUs).

\bibliographystyle{eptcs}
\bibliography{ref,extra}

\end{document}